\documentclass[a4paper,twocolumn]{jpsj2}
\newcommand{\tc}{$T_{\rm c}$}
\newcommand{\msr}{$\mu$SR}
\def\runtitle{}
\def\runauthor{Kazuki \textsc{Ohishi} \textit{et al}.}

\title{Possibility of Horizontal Line Node in KFe$_2$As$_2$ \\
Probed by Muon Spin Rotation}
\author{
Kazuki \textsc{Ohishi}$^{1,6}$\thanks{E-mail:k{\_}ohishi@cross.or.jp, 
Present address: Research Center for Neutron Science and Technology, CROSS, Tokai, Ibaraki 319-1106}, 
Yasuyuki \textsc{Ishii}$^{1,6}$\thanks{Present address: Department of Physics, Tokyo Medical University, 
Shinjuku-ku, Tokyo 160-0023}, 
Hideto \textsc{Fukazawa}$^{2,6}$, Taku \textsc{Saito}$^2$, 
Isao \textsc{Watanabe}$^{1,6}$, Yoh \textsc{Kohori}$^{2,6}$, Takao \textsc{Suzuki}$^{1,3}$, 
Kunihiro \textsc{Kihou}$^{4,6}$, Chul-Ho \textsc{Lee}$^{4,6}$, Kiichi \textsc{Miyazawa}$^{4,5,6}$
Hijiri \textsc{Kito}$^{4,6}$, Akira \textsc{Iyo}$^{4,5,6}$, Hiroshi \textsc{Eisaki}$^{4,6}$
}
\inst{
$^1$Advanced Meson Science Laboratory, Nishina Center for Accelerator-Based Science, 
RIKEN, Wako, Saitama 351-0198
\\
$^2$Department of Physics, Chiba University, Chiba 263-8522
\\
$^3$Graduate School of Arts and Science, International Chirstian University, Mitaka, Tokyo 181-8585
\\
$^4$National Institute of Advanced Industrial Science and Technology (AIST), Tsukuba, Ibaraki 305-8568
\\
$^5$Department of Applied Electronics, Tokyo University of Science, Noda, Chiba 278-8510
\\
$^6$JST, Transformative Research-Project on Iron Pnictides (TRIP), Chiyoda-ku, Tokyo 102-0075
}
\abst{
The anisotropy and temperature dependence of the magnetic field penetration depth in the iron-based 
superconductor KFe$_2$As$_2$ have been measured down to 20 mK by means of muon spin rotation. 
We have observed that the flux-line lattice forms triangular symmetry when we applied the field parallel to $c$-axis, 
suggesting the interaction between vortices should be isotropic. 
The temperature dependence of the penetration depth observd with an applied field 
parallel and perpendicular to $c$-axis are different. 
These results can be accounted for by a superconducting gap function with a horizontal line node in the basal plane.  
}

\kword{iron-based superconductor, penetration depth, Ba$_{1-x}$K$_x$Fe$_2$As$_2$, \msr}

\begin{document}
\maketitle

The discovery of superconductivity at \tc\/ =  26 K in LaFeAsO$_{1-x}$F$_x$ \cite{Kamihara2008} 
and the improvement of \tc\/ above 50 K in other iron pniceides \cite{Kito2008,Ren2008a} have 
stimulated a tremendous research effort to elucidate the superconducting pairing mechanism and 
symmetry in iron pnictide superconductors. In oxygen-free system of $A$Fe$_2$As$_2$ ($A$ = Ca, 
Sr, Ba, and Eu) \cite{Rotter2008,Alireza2009}, the emergence of superconductivity seems to be 
associated with the suppression of magnetic ordering, which is accomplished either via 
application of pressure \cite{Alireza2009,Torikachvili2008,Park2008} 
or by chemical substitutions \cite{Rotter2008,Sefat2008,Jiang2009,Kumar2009,Pratt2009}. 
%
In the optimally doped Ba$_{1-x}$K$_x$Fe$_2$As$_2$ ($x\sim 0.4$) 
with \tc\/~=~38~K, the Fermi surface consists of well-separated 
hole and electron sheets, which seem to exhibit good interband nesting, as found in many iron pnictide 
superconductors. For example, the angle resolved photoemission spectroscopy (ARPES) \cite{Ding2008}, muon 
spin rotation (\msr) \cite{Khasanov2009}, and NMR \cite{Yashima2009} have suggested the multiple full gap 
with sign change between these two different bands, namely $s_\pm$-wave. 
Meanwhile in KFe$_2$As$_2$, which shows superconductivity at \tc\/ = 3.5K, 
de Haas-van Alphen experiment and the accompanying band structure 
calculations carried out by Terashima {\it et al.} \cite{Terashima2010} 
have shown that the electron sheets centered at the X point of a Brillouin zone are replaced with small hole tubes, 
therefore, the interband nesting observed in optimally doped sample is not there anymore. 
This unusual electronic properties was also confirmed by ARPES \cite{Sato2009} and recent neutron measurement \cite{Lee2011}. 
Other studies, i.e., NQR and specific heat \cite{Fukazawa2009}, thermal conductivity \cite{Dong2010}, 
NMR \cite{Zhang2010} and 
penetration depth \cite{Hashimoto2010}, have suggested that the superconducting gap in KFe$_2$As$_2$ has 
line node in contrast to the nodeless gap suggested in Ba$_{0.6}$K$_{0.4}$Fe$_2$As$_2$. 
On the other hand, an isotropic triangular flux-line lattice (FLL) was observed by 
the small angle neutron scattering (SANS) without FLL transitions up to 
$H_{c2}$ with the condition of $H$ parallel to $c$-axis \cite{Furukawa2011}. 
This result suggests that an interaction between vortices is isotropic. 
According to the theoretical approach, it is suggested that horizontal node (node in $ab$-plane) 
appears in the hole band at $\Gamma$ point of a Brillouin zone in KFe$_2$As$_2$ \cite{Suzuki2011}, 
while some closed the nodal loops at the electron Fermi surface was suggested in relative superconductor of 
BaFe$_2$(As$_{1-x}$P$_x$)$_2$ \cite{Yamashita2011}. 
Although various possiblities of line node in KFe$_2$As$_2$ are suggested, 
the reported experimental data are observed in polycyristalline samples or in single crystals 
only with the condition of an applied field $H$ parallel to $c$-axis. 
In order to elucidate the superconducting 
gap structure in KFe$_2$As$_2$, we have observed the temperature dependence of penetration depth for 
both $H\parallel c$ and $H\perp c$-axis down to 20~mK. 

In this Letter, we report the demonstration of the horizontal nodal superconductivity in KFe$_2$As$_2$ 
by means of \msr\/ measurements. Our findings reveal that the temperature dependence of penetration depth 
measured only with one field direction, i.e., $H\parallel c$, cannot confirm the existence of horizontal node, 
and clarify the pairing symmetry in heavily hole doped KFe$_2$As$_2$. 
\begin{figure}[tbp]
\begin{center}
\rotatebox[origin=c]{0}{\includegraphics[width=0.8\columnwidth]{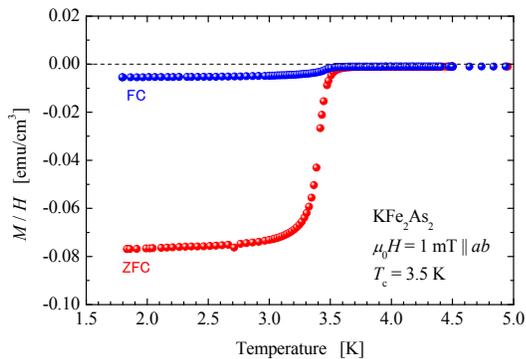}}
\caption{(color online) The dc magnetic susceptibility at $\mu_0H=1$~mT observed after zero field cooled (ZFC) 
and field cooled (FC).}
\label{fig1}
\end{center}
\end{figure}

High quality single crystalline samples of KFe$_2$As$_2$ were grown by a self-flux method 
which will be described in detail elsewhere \cite{Kihou2010}. Figure~\ref{fig1} shows the temperature dependence of 
susceptibility below 5~K observed at $\mu_0H=1$~mT after zero filed cooled and field cooled. 
The \tc\/ was estimated to be 3.5~K. The \msr\/ measurements were conducted both at M15 beamline at 
TRIUMF, Vancouver, Canada and at RIKEN-RAL Muon Facility in the Ruthreford Appleton Laboratory, Didcot, UK. 
The transverse field (TF) \msr\/ measurements were performed at temperatures from 
20 mK to 5 K. In two sets of TF-\msr\/ measurements, the magnetic field was applied in parallel and 
perpendicularly to the crystallographic $c$-axis, respectively, and always perpendicularly to the 
implanted muon spin polarization. The sample was field cooled at the measured magnetic fields to minimize disorder of 
the FLL due to flux pinning. 
Since the muons stop randomly on the length scale of the FLL, the muon spin precession signal provides a random 
sampling of the internal field distribution in the FLL state. 

\begin{figure}[tbp]
\begin{center}
\rotatebox[origin=c]{0}{\includegraphics[width=0.8\columnwidth]{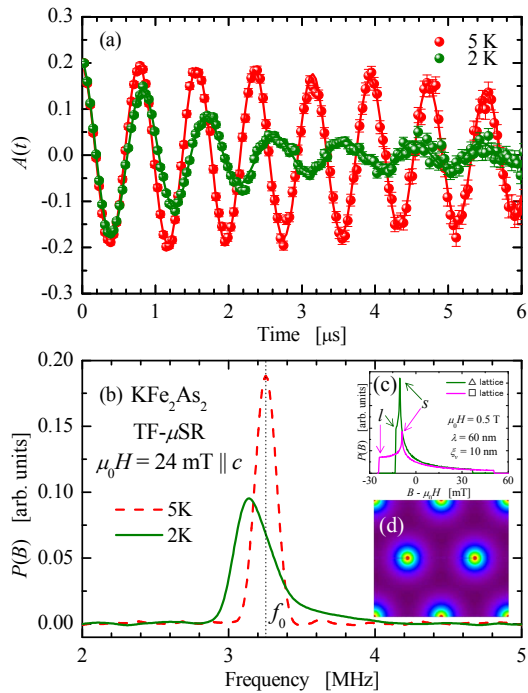}}
\caption{(color online) (a) Muon TF precession signals at normal state (5 K) and 
flux-line lattice (FLL) state (2 K) of single crystalline sample KFe$_2$As$_2$ with $\mu_0H=24$~mT 
parallel to $c$ axis. The solid lines represent the result of fitting analysis (see text). 
(b) Local magnetic field distribution $P(B)$ in the normal and FLL state. 
The dotted line stands for the frequency $f_0$, which corresponds to the applied field. 
(c) Theoretical $P(B)$ distributions ($\lambda=60$~nm, $\xi_v=10$~nm, 
and $\mu_0H=0.5$~T) for triangular FLL (green) and square FLL (pink) state.
(d) The contour map of the FLL at 2~K in real space. }
\label{fig2}
\end{center}
\end{figure}
The time evolution of the muon spin polarization signals measured with $\mu_0H=24$~mT 
parallel to $c$-axis above and below \tc\/ are presented in Fig. \ref{fig2}(a). Under a TF and below \tc\/, 
implanted muons experience an inhomogeneity of the field due to FLL formation 
that leads to relaxation. As shown in Fig.~\ref{fig2}(a), complete depolarization is observed at 2~K, 
indicating that the entire volume falls into the superconducting state. Figure~\ref{fig2}(b) shows 
the magnetic field distributions $P(B)$ obtained by means of the fast Fourier transforms of TF-\msr\/ signals. 
For the superconductor in the FLL state, $P(B)$ is uniquely determined by the magnetic penetration 
depth $\lambda$ and coherence length $\xi$ \cite{Brandt1988}. 
The $P(B)$ at 5~K looks very symmetric around the frequency $f_0$, which corresponds to the applied field, 
because the applied field penetrates into the sample homogeneously. On the other hand, $P(B)$ at 2~K 
becomes asymmetric due to FLL formation. 
Fig.~\ref{fig2}(c) shows the calculation results of magnetic field distribution 
for both triangular and square symmetries of FLL. 
The difference between the low field cutoff ($l$) and saddle-point ($s$) fields is 
larger in a square FLL than in a triangular one. 
Thus, a large shoulder on the low field side of the peak is a signature of a square FLL. 
Comparing the observed and calculation results, it is obvious that the FLL forms triangular symmetry, 
consistent with SANS results \cite{Furukawa2011}.  

The spacial magnetic field distribution in a FLL state is 
\begin{align}
B(\mathbf{r})=B_0\sum_\mathbf{K}\frac{e^{-i\mathbf{K}\cdot\mathbf{r}}e^{-K^2\xi_v^2}}{1+K^2\lambda^2},\\
P(B)=\langle\delta\left(B-B(\mathbf{r})\right)\rangle_\mathbf{r},
\label{distribution}
\end{align}
where $\mathbf{K}$ is the reciprocal lattice vector, $B_0$ is the average internal field, $\mathbf{r}$ is the 
vector coordinate, and $\xi_v$ is the cutoff parameter. 
The TF-\msr\/ time spectra were fitted to a theoretical function by assuming the internal field distribution $P(B)$ 
obtained from Eq.~(\ref{distribution}) and accounting for the FLL disorder and the nuclear moment contributions, 
\begin{equation}
A(t)=Ae^{-(\sigma_p^2+\sigma_n^2)t^2/2}\int P(B)e^{i(\gamma_\mu Bt-\phi)} dB,
\label{timefit}
\end{equation}
where $A$ is the initial asymmetry, 
$\sigma_p$ and $\sigma_n$ denote the relaxation due to the FLL disorder and the nuclear moment, respectively, 
$\gamma_\mu/2\pi=135.5$~MHz/T is the muon gyromagnetic ratio and $\phi$ is the initial phase. 
Fig.~\ref{fig2}(d) shows the contour map of a FLL at 2~K obtained by the fitting analysis. 

\begin{figure}[tbp]
\begin{center}
\rotatebox[origin=c]{0}{\includegraphics[width=0.8\columnwidth]{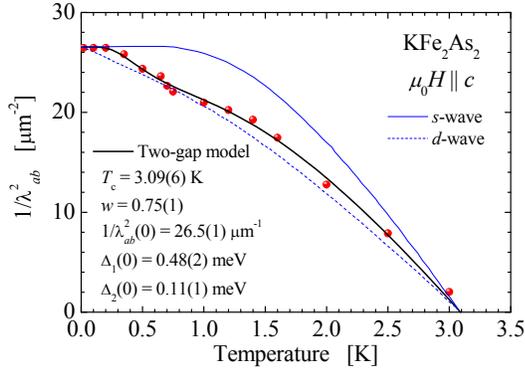}}
\caption{(color online) Temperature dependence of the inverse squared magnetic penetration depth 
$1/\lambda_{ab}^2$ obtained with $\mu_0H=24$~mT parallel to $c$-axis. The solid black line represents the result of 
a fitting analysis with the two-full-gap model. For comparison, the calculated lines for clean $s$-wave and $d$-wave 
pairings are also presented.}
\label{fig3}
\end{center}
\end{figure}

Figure~\ref{fig3} shows the temperature dependence of the inverse squared magnetic penetration depth 
$1/\lambda^2_{ab}(T)$. A dip was observed around 0.75~K, similar to the results in Ba$_{1-x}$K$_x$Fe$_2$As$_2$ 
\cite{Khasanov2009,Goko2009,Ohishi}, which are well described by the multiple full-gap behavior. 
As shown in Fig.~\ref{fig3}, both single $s$- and $d$-wave curves show significant departure from the experimental data. 
The experimental data were analyzed within the framework of the phenomenological $\alpha$ model by assuming two 
independent contributions to $\lambda^{-2}$ \cite{Khasanov2007,Carrington2003},
\begin{equation}
\frac{\lambda^{-2}(T,\Delta)}{\lambda^{-2}(0)}=w\frac{\lambda^{-2}(T,\Delta_1)}{\lambda^{-2}(0)}+(1-w)\frac{\lambda^{-2}(T,\Delta_2)}{\lambda^{-2}(0)},\label{tdep1}
\end{equation}
\begin{equation}
\frac{\lambda^{-2}(T,\Delta_i)}{\lambda^{-2}(0)}=1+\frac2{\pi}\int_0^\pi\!\!\!\int_{\Delta_i(T,\theta)}^\infty\!\!\!\!\frac{\partial f(E)}{\partial E}\frac{E\sin{\theta}dEd\theta}{\sqrt{E^2-\Delta_i(T,\theta)^2}}, \label{tdep2}
\end{equation}
where 
$w$ is the ratio of gap energy between the two gaps, $f(E)$ is the Fermi function, $\Delta_i(T,\theta)$ is the 
superconducting gap; for simplicity, the $\Delta_i(T,\theta)=\Delta_i(T)\cos{\theta}$ for the nodal gap and 
$\Delta_i(T)$ for the full gap. The temperature dependence of the gap is approximated by 
$\Delta_i(T)=\Delta(0)\tanh\{{1.82[1.018(T_{\rm c}/T-1)]^{0.51}}\}$ \cite{Carrington2003}. 
The result of analysis with full gap  model is presented in Fig.~\ref{fig3} by solid black line, 
yielding $1/\lambda_{ab}^2(0)=26.5(1)$~$\mu$m$^{-2}$, $w=0.75(1)$, $\Delta_1(0)=0.48(2)$~meV, 
$\Delta_2(0)=0.11(1)$~meV. 

\begin{figure}[tbp]
\begin{center}
\rotatebox[origin=c]{0}{\includegraphics[width=0.8\columnwidth]{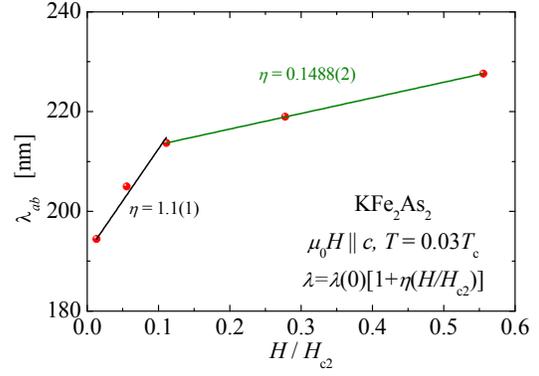}}
\caption{(color online) Magnetic field dependence of $\lambda_{ab}$ obtained at $T=0.03T_{\rm c}$ 
with various applied fields. The solid lines show the result of fitting analysis (see text). The $H_{c2}$ at this temperature 
is approximately 1.8~T.}
\label{fig4}
\end{center}
\end{figure}

The field dependence of $\lambda_{ab}$ observed at $T=0.03T_{\rm c}$ is shown in Fig.~\ref{fig4}. 
It clearly exhibits two kinds of field dependence, where the gradient changes at around $H/H_{c2} = 0.1$. 
The increase of $\lambda_{ab}$ is attributed to the anisotropic order parameters because it is supposed to be a 
constant in the conventional $s$-wave superconductors. 
A fitting by the relation $\lambda(H/H_{\rm c2}) = \lambda(0)[1+\eta\cdot (H/H_{\rm c2})]$ provides 
a dimensionless parameter $\eta$ which represents the strength of the pair breaking effect. 
From the analysis, we find that $\eta=1.1(1)$ and $\eta=0.1488(2)$ for $H/H_{c2}\le 0.1$ and 
$H/H_{c2}\ge 0.1$, respectively. 
The large value of $\eta$ for $H/H_{c2}\le 0.1$ is attributed to be quasiparticle excitations 
from the smaller gap $\Delta_2$ due to the effect of Doppler shift. 
On the other hand, the $\eta$ is small above $H/H_{c2} = 0.1$ 
because the larger gap $\Delta_1$ is robust against the field.

\begin{figure}[tbp]
\begin{center}
\rotatebox[origin=c]{0}{\includegraphics[width=0.8\columnwidth]{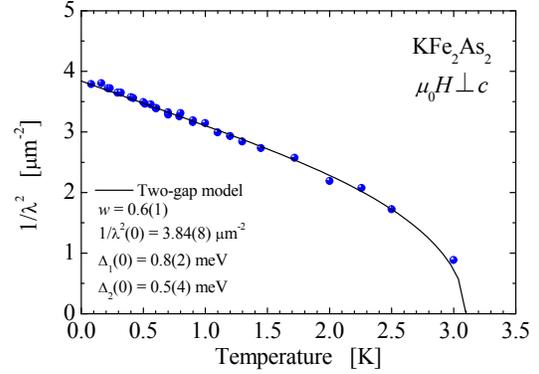}}
\caption{(color online) Temperature dependence of the inverse squared magnetic penetration depth 
$1/\lambda^2$ obtained with $\mu_0H=16$~mT perpendicular to $c$-axis. 
The solid and dashed lines represent the result of fitting analysis with the line node gap model by two-gap and 
single gap, respectively. }
\label{fig5}
\end{center}
\end{figure}

the data with $H\perp c$ were analyzed using the following function,
\begin{equation}
A(t)=A_se^{-(\sigma_v^2+\sigma_n^2)t^2/2}\cos{(\gamma_\mu Bt+\phi)}
\end{equation}
where  $A_s$ is the initial asymmetry and the depolarization rates $\sigma_v$ is 
the damping due to the FLL formation. 
The superconducting part of the Gaussian depolarization rate $\sigma_v$ can be converted into 
$\lambda$ by \cite{Brandt1988}
\begin{equation}
\sigma_v^2/\gamma_\mu^2=0.00371\Phi_0^2\lambda^4,
\label{lambda}
\end{equation}
where $\Phi_0$ is the magnetic flux quantum. Figure~\ref{fig5} shows $\lambda^{-2}(T)$ obtained from the 
measured $\sigma_v(T)$ by means of Eq.~(\ref{lambda}). 
It is obvious that $T$-linear behavior was clearly observed below $\sim T_{\rm c}/2$, 
suggesting the existence of line node in the superconducting gap. 
The curve fit assuming a \tc\/ = 3.1K, which is obtained by the fitting of 1/$\lambda_{ab}^2$, 
using eq.~(\ref{tdep1}) and (\ref{tdep2}) with $\Delta(T,\theta)=\Delta(T)\cos\theta$ 
perfectly reproduced the data in Fig.~\ref{fig5} 
with $1/\lambda^2=3.84(8)$~$\mu$m$^{-2}$, $w=0.6(1)$, $\Delta_1(0)=0.8(2)$ meV, and $\Delta_2(0)=0.5(4)$ meV.  

We now discuss possible gap structure in KFe$_2$As$_2$. 
Regarding the results obtained with $H\parallel c$, both temperature and field dependence of $\lambda$ indicate 
the two-full-gap or anisotropic gaps. 
Because the increase of $\lambda$ against field is attributed to the anisotropic order parameters and the associated 
nonlinear effect due to the Doppler shift of the quasiparticles in the nodal region ($\Delta(k)\simeq0$) \cite{Volovik1993}. 
It is predicted that $\eta\ll 1$ for the isotropic $s$-wave pairing because the finite gap prevents the shifted 
levels of quasiparticle excitations from being occupied at low temperatures. 
Therefore, the finite value of $\eta$ means the existence of quasiparticle excitations. 
The field dependence of $\lambda$ is expected to be stronger when the phase space satisfying 
$\Delta(k)\simeq0$ has larger volume \cite{Amin2000}, so that 
the quasiparticle excitation observed below $H/H_{c2}=0.1$ suggests the collapse of small gap by the field. 
The $\eta$ becomes small above $H/H_{c2}=0.1$, 
because the quasiparticle excitations are suppressed, suggesting the larger gap is robust against the field. 
The value $\eta=1.1(1)$ is similar to the value of $\eta=0.95(1)$ observed in YNi$_2$B$_2$C \cite{Ohishi2002}, 
which is a superconducotr having point node in the gap \cite{Izawa2002}. 
On the other hand, temperature dependence of $\lambda^{-2}$ with $H\perp c$ shows clear $T$-linear behavior, 
suggesting the existence of line node in the gap. This is not consistent with the results obtained with $H\parallel c$. 

According to the theoretical calculation of the temperature dependence of the superfluid density $n_s$ 
($1/\lambda^2\propto n_s$), 
it is expected that the temperature dependence is different from the relation between the direction of line 
node and that of applied field. For example, it behaves $T$-linear and $T^3$ for the line node perpendicular 
to the field  and that parallel to the field, respectively \cite{Gross1986}. 
These temperature dependence was observed experimentally in B phase of UPt$_3$ \cite{Broholm1990}, 
in which temperature dependence of $\lambda^{-2}$ parallel and perpendicular to $c$-axis behaves different. 
Actually, UPt$_3$ is expected to have not only line node but both line node and point node, but 
anisotropic behavior was observed. 
Thus, considering the result of UPt$_3$, it is obvious that the temperature dependence of $\lambda^{-2}$ 
in KFe$_2$As$_2$ indicate the existence of line node. 
To explain the observed results without any discrepancy, the existence of horizontal line node 
in basal plane is expected. This interpretation is perfectly consistent with other experimental results, 
i.e., NQR and specific heat \cite{Fukazawa2009}, 
thermal conductivity \cite{Dong2010} and penetration depth \cite{Hashimoto2010}. 
Regarding the FLL symmetry observed with $H\parallel c$, the interaction between vortices should be isotropic 
if a line node exists in the basal plane. 
Moreover, the energy scale of the obtained small gap $\sim0.11$ meV $\simeq$ 1.9 T is one order 
larger than the applied field of $H/H_{c2}=0.1$ ($H\sim 0.2$ T). 
The large $\eta$ value obtained below $H/H_{c2}=0.1$ despite a small applied field also suppose the existence 
of line node. 

In summary, we have performed the first measurements of the anisotropic magnetic penetration 
depth $\lambda$ in the iron-based superconductor KFe$_2$As$_2$. 
The $\lambda^{-2}$ decreases proportionally with $T$ when measured 
perpendicular to $c$-axis but with a higher power of $T$ ($\sim T^4$) parallel to $c$-axis. 
Considering the theoretical calculation, the data can be understood by a superconducting gap function 
with a horizontal line node in the basal plane. 

We thank the staff of TRIUMF's Center for Molecular and Materials Science for technical assistance. 
K.O. thanks K. Kubo and H. Sakai for the helpful discussions. 

\end{document}